# The probabilities of an outcome on intervention and control can be estimated by randomizing subjects to different testing strategies – required for assessing diagnostic tests, test trace and isolation and for natural randomisation

Huw Llewelyn MD FRCP
Department of Mathematics
Aberystwyth University
Penglais
Aberystwyth
SY23 3BZ
Tel 01970622802
Fax: 01970622826

hul2@aber.ac.uk

# Randomizing to different diagnostic tests can predict treatment efficacy

Abstract

The efficacy of an intervention can be assessed by randomising patients to different diagnostic tests instead of directly to an intervention and control. This principle is applied by allocating individuals to intervention if the test result is 'positive' (or on one side of a threshold) but allocating individuals to a control if the result is 'negative' (or on the other side of the threshold). This can also be done with different dichotomising thresholds for one test. The frequencies of the outcome in those with each of the four resulting observations are then used to calculate the risk ratio (RR) for the marginal probabilities by solving simultaneous equations. This assumes that the RR due to intervention compared to control is the same in both test groups created by randomisation. The calculations are illustrated by using data from a randomized controlled trial (RCT) that assessed the efficacy of an angiotensin receptor blocker (ARB) in lowering the risk of diabetic nephropathy in patients conditional on urinary albumin excretion rates (AERs). The calculations are also illustrated with simulated data for assessing the effectiveness of test, trace and isolation to reduce transmission of the SARS-Cov-2 virus by randomising to RT-PCR or LFD tests. This approach allows the probabilities of outcomes, their RRs and odds ratios (OR) conditional on the results of covariates (e.g. the AER and RT-PCR test) to be determined, also suggesting a way forward for natural as opposed to active randomisation.

# Randomizing to different diagnostic tests can predict treatment efficacy

## 1. Introduction

It is often not possible to randomize patients directly to intervention or control in clinical trials. This may happen when we wish to assess or to compare the performance of diagnostic tests for predicting response to a treatment or placebo when the latter's efficacy has been established already in a previous RCT. Such tests may have been invented by clinical scientists, artificial intelligence researchers, mathematical modellers and medical statisticians. It is important to assess the performance of such diagnostic tests in order to minimise the risk of failing to treat those who might benefit or to avoid giving treatments with possible adverse effects to those with little chance of benefit. Making the latter error has become known as 'over-treatment'. 'Over-diagnosis' is another concern when a diagnostic label is attached to patients when there is little or no prospect of many patients benefiting from any of the treatments suggested by the label [1].

The variation in response to treatment in patients with different features is also known as the heterogeneity of treatment effect (HTE). This can be tackled by using regression based approaches to predictive heterogeneity of treatment effect analysis, including analyses based on risk modelling (such as stratifying trial populations by their risk of the primary outcome or their risk of serious treatment-related harms) and analysis based on effect modelling (which incorporates modifiers of relative effect) [2, 3]. However, the risk ratio due to a treatment for high blood pressure (BP), for example, will not reduce the overall risk added to by poor diabetic control as treatment for the high BP will not also improve the diabetic control. The estimated risks arising from these models could therefore be assessed in fresh studies to see how well they predict outcomes on individual treatments and controls. This gives rise to the same ethical issues as with single tests if efficacy has already been established in previous RCTs. In order to avoid the ethical issues of repeating RCTs, regression discontinuity design (RDD) might be used as an alternative. This is done by allocating patients to a treatment limb if the result of a test that predicts the outcome is on one side of a threshold and allocating them to a control limb if they are the other side of the threshold. An estimate of risk ratio (RR) or odds ratio (OR) is obtained at the point of discontinuity by assuming that the RR is similar or the same for a result just above or just below the threshold but this approach presents many technical problems [5, 6].

Pearl has pointed out the need for a logical framework for alternative approaches to RCTs of the kind described here based on concepts of causality, counterfactuals and collapsibility [7, 8]. Another approach to assessing how different findings predict outcomes with and without treatment might be to allocate subjects to two different diagnostic testing strategies. This approach is based on a traditional clinical view that a treatment will be more effective if given to patients based on the result of appropriate diagnostic information than if it given to those based on the result of inappropriate information. For example, if an inhaler is given to those with breathlessness and wheeze suggestive of asthma, then more will benefit than when the inhaler is given to those with breathlessness and audible crackles at the lung bases suggestive of left ventricular failure. If the inhaler is truly ineffective, no one will benefit from an inhaler whether they have wheeze (i.e. asthma) or crackles (i.e. heart failure). This principle suggests that the efficacy of a treatment could be assessed by randomizing patients to different diagnostic strategies instead of to a treatment and control, when the latter is difficult to justify ethically.

## 2. Methods of modeling the link between diagnostic tests and treatment efficacy

The aim is to allow the outcome of a trial based on randomising to intervention or control to be predicted by randomising to different diagnostic testing strategies instead. The tests must have different predictive characteristics such as different sensitivities with respect to the outcome. The intervention is applied to a patient if the test result is on one side of a threshold or (when a test is positive) and to a control intervention if it is on the other side of the threshold (or if the same test is negative). This can be done for a pair of different tests or for one test with different thresholds of its numerical test results.

### 2.1 Rationale for methods

Consider that subjects are randomized to take part in two different randomised control trials, Trial 1 and Trial 2 as shown in Figure 1. In Trial 1, the test T1 is performed on all subjects before they are randomised again a second time into those to be given a control or intervention. In those randomized to Trial 2, a test T2 is performed before randomisation again to control or intervention. The first assumption (A) is that the risk ratio due to the intervention in both trials T1 and T2 is due to the direct effect of the intervention on the outcome (with no causal effect from Test T1 and T2 on the outcome). The second assumption (B) is that the frequency of occurrence of the results of test T1 and the results of test T2 in those with the outcome on treatment is the same in as in those with the outcome on control. It then follows that if the proportion with an outcome on control in those who test T1 negative is a, then the reduced risk with intervention is a*r. Similarly if the proportion with an outcome on intervention in those testing T1 positive is b, then the increased risk on control is b/r. The same applies in Trial 2 when the outcomes are proportions c, c*r, d and d/r.

Figure 1: Diagram of randomisation to control or intervention in two trials after testing with test T1 in Trial 1 and testing with test T2 in Trial 2.

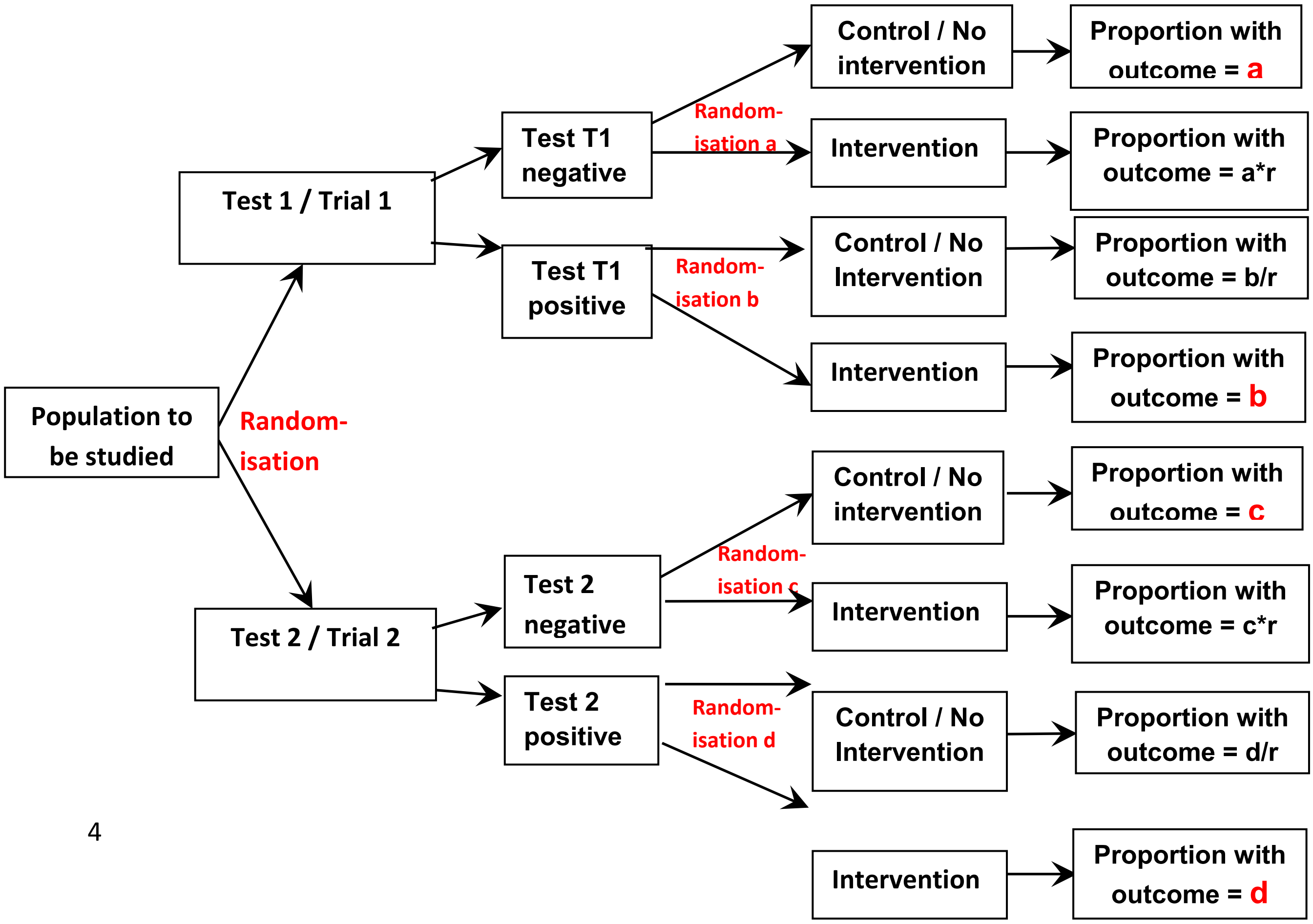

2.2 Initial randomisation to different testing strategies

We now perform a different study design as shown in Figure 2. We again randomize subjects to two groups, testing one group with test T1 and the other with test T2. However instead of randomizing again to control or intervention, we allocate subjects to a control if their test is negative and to intervention if the test is positive. In this design there are only 4 observed proportions, a. b. c and d as shown in Figure 2. However, these are the same proportions a, b, c and d shown in Figure 1. The risk ratio of r is the same in Figures 1 and 2 also.

Figure 2: Diagram of randomisation to different tests and allocation to control if a test is negative or to intervention if the test is positive

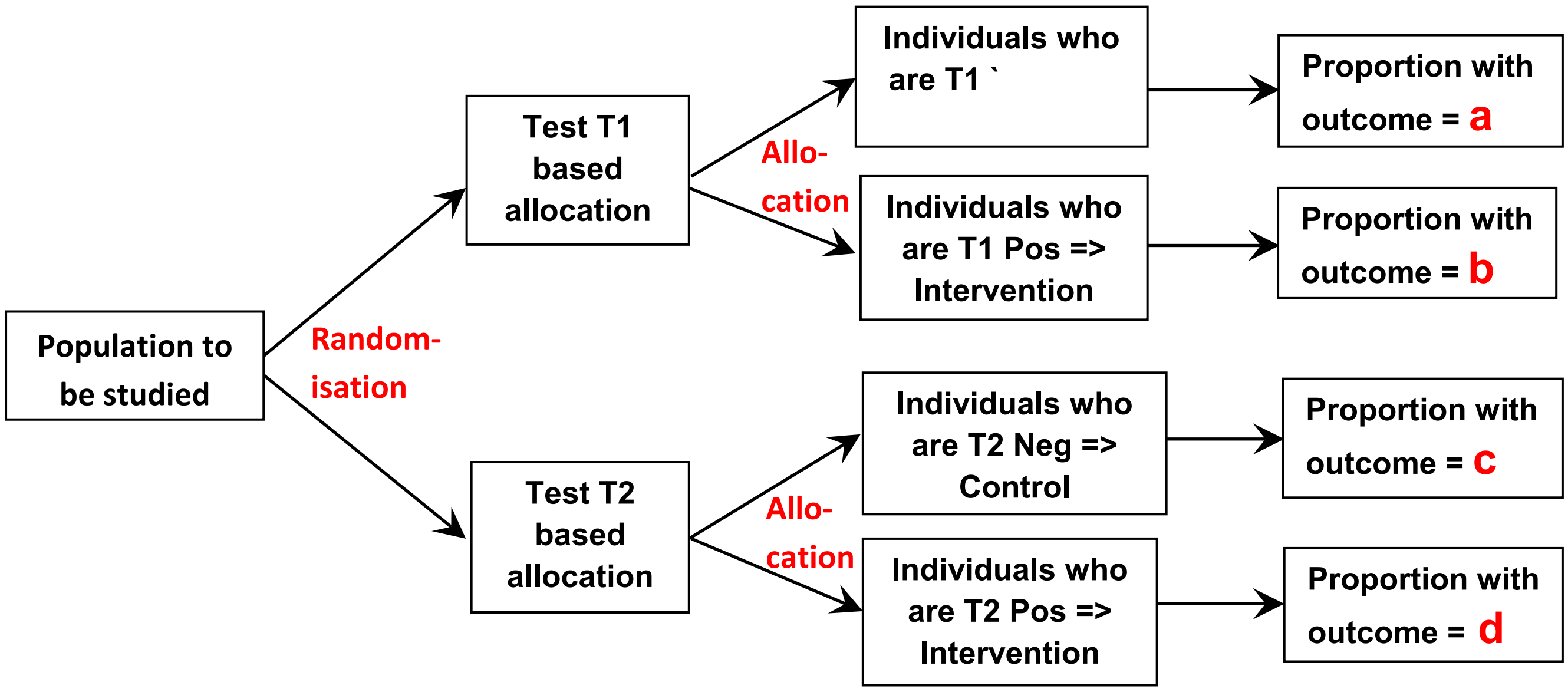

2.3 Proportions with various outcomes

In Figure 2 the proportion a = the observed overall proportion with the adverse outcome and also having had a NEGATIVE result of test T1 and thus having been allocated to a CONTROL (see top line of Figure 2):

Again in Figure 2, r = is the risk ratio so that a*r = the calculated UNOBSERVED proportion having the adverse outcome and also having a NEGATIVE result of test T1 and thus having been allocated to the INTERVENTION (therefore calculated from knowing 'a' and r)

The proportion b = the observed proportion with the adverse outcome and also having had a POSITIVE result of test T1 and thus having been allocated to the INTERVENTION

The proportion b/r = the calculated UNOBSERVED proportion having the adverse outcome, also having a POSITIVE result of test T1 and having been allocated to the CONTROL (therefore calculated from knowing 'b' and r)

The proportion c = the observed proportion with the adverse outcome and also having had a NEGATIVE result of test T2 and thus having been allocated to a CONTROL

The proportion c*r = the calculated UNOBSERVED proportion with the adverse outcome, also having a NEGATIVE result of test T2 and having been allocated to the INTERVENTION (therefore calculated from knowing c and r)

The proportion d = the observed proportion with the adverse outcome and also having had a POSITIVE result of test T2 and thus having been allocated to the INTERVENTION

The proportion d/r = the calculated UNOBSERVED proportion having the adverse outcome, also having a POSITIVE result of test T2 and having been allocated to the CONTROL (therefore calculated from knowing d and r)

2.4 Simultaneous equations

Let a + a*r +b/r + b = y, the probability of having the outcome when randomly allocated to Test 1

Let c + c*r +d/r + d = y, the probability of having the outcome when randomly allocated to Test 2

As the overall prior probability 'y' of having the outcome is the same in the groups randomly allocated to test T1 and T2:

$a + a*r + b/r + b = y = c + c*r + d/r + d$ Equation 2.1

Omitting y and rearranging Equation 1: $a*r - c*r + b/r - d*r = c + d - a - b$ Equation 2.2

Rearranging Equation 2: $r^2(a-c) - r(c + d - a - b) - (b-d) = 0$ Equation 2.3

Rearranging Equation 3: $(a-c)r^2 + (a-c)r + (b-d)r - (b-d) = 0$ Equation 2.4

Factorising Equation 4: $((a-c)r + (b-d))(r-1) = 0$ Equation 2.5

From Equation 5 either: $(r+1) = 0$ and $r = -(b-d)/(a-c) = (d-b)/(a-c)$ Equation 2.6

… or $-(b-d)/(a-c) = 0$ and $r = 1$ Equation 2.7

Therefore $r = -(b-d)/(a-c) = (d-b)/(a-c)$ = the risk ratio ratio Equation 2.8

For example, when a = 0.028, b = 0.003, c = 0.016 and d = 0.006, then

Risk ratio is: $r = (d-b)/(a-c) = (0.006-0.003)/(0.028-0.016) = 0.25$ Equation 2.9

The probability of the outcome conditional on T1 or T2 is: $y = a + a*r + b/r + b = c + c*r + d/r + d =$
$= 0.028+0.007+0.012+0.003 = 0.016 + 0.004+0.024+0.06 = 0.05$ Equation 2.10

3. Results based on real and simulated examples

3.1 Example based on real data

The following illustrative example is based on the result of a randomised controlled trial comparing the effect of placebo and irbesartan on the proportion of Type 2 diabetic patients who develop 'Nephropathy' in the form of severe proteinuria with an albumin exertion rate (AER) of over 200mcg/min within 2 years [9]. This AER range of >200mcg/min is regarded as one of the sufficient diagnostic criteria for the diagnosis of 'Nephropathy'. This diagnosis suggests that the patient is in danger of suffering progressive renal impairment perhaps requiring renal dialysis and other support. The term ‘Nephropathy’ is also be used to indicate severe proteinuria within 2 years. The predicting test used was also the albumin excretion rate (AER) performed at the beginning of the trial. Note that randomisation was to 3 limbs. For the sake of simplicity the two intervention limbs are combined. The data in Table 1 show that the proportion developing nephropathy after 2 years on placebo was 30/196. However, the proportion developing nephropathy after 2 years on either dose of irbesartan was 29/379. This means that the risk ratio was (29/379)/(30/196) = 0.499.

The pair of dichotomous test results T1 and T2 can be different tests such as a RT-PCR and Lateral Flow Device (LFD) or different dichotomising thresholds of a single numerical test such as an AER. This illustration will be based on thresholds of an AER of 40mcg/min and an AER of 80mcg/min. Thus a T1 positive was an AER >80mcg/min and T1 negative was an AER ≤ 80mcg/min. A T2 positive was an AER >40mcg/min and T2 negative was an AER ≤ 40mcg/min. If patients were randomised to T1 then the AER threshold would be 80mcg/min and if randomised to T2, the AER threshold would be 40mcg/min. Note that the results in shaded data in Table 1 would not have been seen by using this strategy.

**Table 1** Proportion of patients developing nephropathy up to 24 months on different interventions after starting from different baseline urinary albumin excretion rates (AERs)

| Baseline AER | Placebo | Irbesartan 150mg od | Irbesartan 300mg od |
|---|---|---|---|
| 161 to 200 µg/minute | 2/7 = 28.57% | 4/13 = 30.77% | 1/2 = 50.00% |
| 121 to 160 µg/minute | 9/23 = 39.13% | 3/16 = 18.75% | 0/11 = 0.00%* |
| 81 to 120 µg/minute | 9/32 = 28.13% | 7/33 = 21.12% | 4/37 = 10.81% |
| 41 to 80 µg/minute | 9/57 = 15.79% | 5/66 = 7.58% | 4/74 = 5.41%† |
| 20 to 40 µg/minute | 1/77 = 1.30% | 0/59 = 0% | 1/68 = 1.47% |
| | | | |
| All: 20 to 200µg/minute | 30/196 = 15.30% | 19/187 = 10.16% | 10/192 = 5.21%# |
| **Risk ratio for placebo and both doses of irbesartan = (29/379)/(30/196) = 0.499** | | | |

The number of patients with an AER ≤40mcg/min allocated to placebo in Table 1 is 77. The number of patients with an AER>40mcg/min and allocated to treatment in Table 1 was 66 +74 + 33+ 37 + 16 + 11 + 13 + 2 = 252, which was 252/2 = 126 per limb. Therefore without having all the data in Table 1 available except for the un-shaded area, the estimated total number of patients in each limb is 77+126 = 203. This means that an estimated 203 patients were allocated to placebo and 406 were allocated to treatment with either dose of irbesartan. By performing the same exercise based on an AER threshold of 80mcg/min. the number of patients randomised to placebo <80mcg/min was 77 + 57 = 134. The number of patients allocated to treatment with an AER >80mcg/min was 33+ 37 + 16 + 11 + 13 + 2 = 112 or 112/2 = 61 patients per limb. The estimated total number of patients allocated to each limb based on a threshold of AER = 80mcg/min is therefore 134+ 61 = 195.  The average of these two estimates is (195 + 203)/2 = 199 per limb. This means that the estimated number randomised to placebo was 199 and to treatment was 199 x 2 = 398.

3.2 Calculating estimates of the risk ratio and unobserved proportions

From Table 1, the estimated proportion of the outcome of nephropathy and having an AER ≤80mcg/min on placebo is 10/199 so that the estimated probability is 10/199 = 0.0503. This corresponds to probability 'a' in the above rationale. The estimated proportion of the outcome of nephropathy and having an AER >80mcg/min on treatment is 10/398 so that the estimated probability is 10/398 = 0.0477. This corresponds to probability 'b' in the above rationale. The estimated proportion of the outcome of nephropathy and having an AER ≤40mcg/min on placebo is 1/199 so that the estimated probability is 1/199 = 0.0050. This corresponds to probability 'c' in the above rationale. The estimated proportion of the outcome of nephropathy and having an AER >40mcg/min on treatment is 28/398 so that the estimated probability is 28/398 = 0.0704. This corresponds to probability 'd'.

We are now in a position to calculate the estimated risk ratio ratio. The probability a = 10/199 = 0.0503, b= 19/398 = 0.0477, c = 1/199= 0.0050 and d= 28/398 = 0.0704. The calculated estimated risk ratio is thus r = (d-b)/(a-c) = (28/398-19/398)/(10/199-1/199) = (9/398)/(9/199) = 0.5. This allows us to calculate the estimated unobserved proportions of nephropathy in those on treatment and control as shown in Table 2.

The proportion developing nephropathy on treatment and an AER≤80mcg/min is 10/199*0.5 = 10/398 = 0.0251. The calculated estimated proportion developing nephropathy on treatment and an AER≤40mcg/min is 1/199*0.5 = 1/398 = 0.0025. The calculated estimated proportion developing nephropathy on control and an AER>80mcg/min is (19/398)/0.5 = 19/199 = 0.948. The proportion developing nephropathy on control and an AER>40mcg/min is (29/398)/0.5 = 29/199 = 0.1408

The estimated observed and unobserved probabilities of nephropathy in those on treatment and control are shown in the upper row of Table 2. The estimated total proportion developing nephropathy on control in the top row is 10/199 + 19/199 = 29/199. The estimated total proportion developing nephropathy on treatment in the top row is also 10/398 + 19/398 = 29/398.

Table 2: Estimated observed and unobserved probabilities of nephropathy in those on treatment and control

T1: Threshold of AER = 80mcg/min.

| | | | |
|---|---|---|---|
| AER≤80mcg/min<br>Control (a)<br>(Observed)<br>a=10/199=0.0503 | *AER≤80mcg/min*<br>*Treatment (a*r)*<br>*(Calculated)*<br>*a*r=(10/199)*0.5= 10/398 = 0.0251* | *AER>80mcg/min*<br>*Control (b/r)*<br>*(Calculated)*<br>*b/r=(19/398)/0.5= 19/199 = 0.0954* | AER>80mcg/min<br>Treatment (b)<br>(Observed)<br>b=19/398 = 0.0477 |

T2: Threshold of AER = 40mcg/min

| | | | |
|---|---|---|---|
| AER≤40mcg/min<br>Control (c)<br>(Observed)<br>c=1/199=0.0050 | *AER≤40mcg/min*<br>*Treatment (c*r)*<br>*(Calculated)*<br>*c*r=(1/199)*0.5= 1/398 = 0.0025* | *AER>40mcg/min*<br>*Control (d/r)*<br>*(Calculated)*<br>*d/r=(28/398)/0.5= 28/199 = 0.1408* | AER>40mcg/min<br>Treatment (d)<br>(Observed)<br>d=28/398=0.0704 |

| |
|---|
| Risk ratio= r = (d-b)/(a-c) = (28/398-19/398)/(10/199-1/199) = (9/398)/(9/199) = 0.5 |

3.3 Some stochastic and other issues

The risk ratio from Table 1 was (29/379)/(30/196) = 0.499. The calculations in Table 2 give an estimate of (9/398)/(9/199) = 0.5 which happens to be identical to that using all the data in Table 1. These identical results are clearly fortuitous in view of the small numerators of 9 in each case. The calculations summarised in Table 2 are estimating the result of an RCT with 196 subjects in the placebo limb and 379 subjects in the Irbesartan limb where the outcome was nephropathy AND a baseline AER between 40 and 80mcg/min. When the observed proportions are 9/379 and 9/196 the P value for the difference is 0.064. However when the observed proportions are 29/379 and 30/196 the P value for the difference is 0.002. In order to achieve the same P value for the range 40 to 80mcg/min, about 3.6 times as many subjects would have to be recruited into the trial if the same proportions prevailed.

If there had been very large numbers of subjects, then the risk ratios of nephropathy AND an AER in the other ranges would be expected to be the same. In the AER range 20 to 40mcg/min in Table 1 the risk ratio point estimate was the same again at 1/379 and 1/196 = 0.5. However, for an AER between 80 and 200mcg/min the proportions are 19/379 and 20/196 giving a risk ratio of 0.491, this being in keeping with the assumption that measuring the AER as no causal effect on the outcome of nephropathy. In order to conduct such a study subjects would have to be randomised into the 3 potential limbs of Placebo, Irbesartan 150mg or Irbesartan 300 mg but the medication would only be administered if the patient baseline AER were between 40 to 80mcg/min where it were considered that there was equipoise.

Subjects with baseline AER below 40mcg/min might be allocated to placebo and those with a baseline AER above 80mcg/min allocated to a treatment in order to construct curves that showed the probability of developing nephropathy on control and treatment for all baseline AERs from 20 to 200mcg/min. [4]. However although the strategy would be explained to subjects they would have to be ‘blinded’ to the result of their baseline AER and the subsequent nature of what was administered. In order to get sufficient statistical power and meaningful differences, the numbers randomised would have to be very large. This approach might be of value when monitoring the efficacy of treatments during a day to day audit care and assessing newer diagnostic tests (e.g. to assess the performance of the the simpler albumin creatinine ratio as a possible replacement for the AER).

These point estimates from the overall proportions developing nephropathy from using all the data in Table 1 were 30/196 = 0.153, and on treatment they were 29/379 = 0.0765. However from randomising to different diagnostic strategies the estimated overall proportion with nephropathy on control was 29/199 = 0.1457, and on treatment it was 29/398 = 0.0729. Clearly, precise results can be established only with a very large or infinite number of observations. However, the simplicity of randomising to different diagnostic tests instead of treatments means that it should be easier to recruit larger number of subjects that would reduce the width of the confidence intervals. The object of this paper is to demonstrate the principle of the approach. Placebo would be given to lower risk patients at lower risk of an adverse outcome and treatment given to those at higher risk. This might also be an advantage when it comes to assessing the effectiveness of a diagnostic and treatment strategy where randomisation of subjects to treatment or control would be problematic (e.g. during ‘test, trace and isolation’ (TT&I) for Covid-19).

3.4 Natural randomisation

A clinical guideline might be proposed as a result of the above findings, that suggests that patients with an AER below 40mgc/min should not be treated as the difference in the risk of nephropathy with and without treatment was too small to warrant the risk of side effects from the medication. Similarly the guideline might suggest treatment might be given if the AER were above 80mcg/min. If the AER were between 40 and 80mcg/min then guideline users might be asked to apply their own discretion. This might result in most patients with an AER just above 40mcg/min not being treated but most of those with an AER just below 80mcg/min being treated. At some value of AER, an equal number of patients would be treated and not treated in an apparent random manner due to uncertainty. The patients at this point of maximum uncertainty and perhaps those just above and just below this point of uncertainty could be regarded as the subjects of a natural experiment. The number of patients who develop nephropathy in those allocated to treatment could be compared with the number with nephropathy not treated to provide a risk ratio as described in Section 3.3. Other information arising from applying the guideline would be the proportions of those with an AER above and below the point of uncertainty that develop nephropathy and also the distribution of AER values in all those to whom the guideline is applied. Under ideal conditions, this would allow the result of the RCT that gave rise to the guideline to be replicated in the 'real world'. This would be an interesting hypothesis to test with an observational study for any treatment. If successful it would be a more convenient way for comparing the predictive performance of different diagnostic tests such as comparing the albumin creatinine ratio with the AER by avoiding randomising into different diagnostic strategies.

In terms of the vocabulary of causal inference, the decision under uncertainty of the patient and or doctor is an instrumental variable. It shows exclusivity by not affecting the outcome of nephropathy directly but only doing so via the decision to treat or not to treat, thus satisfying the condition of relevance. Finally the condition of random assignment is met by the equipoise and uncertainty experienced by the decision makers at one point resulting in them allocating in the long run an equal number of patients at random to treatment and control.

4.1 Applications to TT&I for Covid-19 using simulated data from a suggested study design

Table 3 shows some simulated results from a suggested cluster design where people from different communities (e.g. schools) are randomised into 3 groups: (1) the RT-PCR group, (2) the LFD group with delay and (3) the LFD group with no delay. In Group 1, subjects testing positive for RT-PCR are asked to isolate 48 hours from when the test was performed (to ensure that all results were back) and those testing negative are asked not to isolate. In Group 2, those testing positive for a LFD test are asked to isolate 48 hours from when the test was performed (so that isolation was started after the same delay as for the RT-PCR group) but those with negative results are asked not to isolate. In Group 3, isolation is started immediately that LFD positive result becomes available (e.g. after 30 minutes). Both PRT-PCR and LFD tests are performed on all participants in the 3 groups as baseline. However, the decision in group 1 is based on the RT-PCR result and the decision in groups 2 and 3 is based solely on the LFD test result.

All participants in both groups testing positive and negative at day zero are asked to keep a record of contacts within two metres for more than 15 minutes for the next fixed number of days (perhaps with a smart-phone app). After the fixed number of days (e.g. between 5 and 10) all the contacts of the 3 groups are tested with RT-PCR and LFD and those in the group who were tested negative originally but converted to be tested positive with either test at the fixed number of days later are designated ‘infected contacts’ and are ‘backward traced’ [10]. If they had been in contact within 2 metres for more than 15 minutes with a subject testing positive at the outset, the latter is designated a ‘positive spreader’ and the newly infected individuals termed ‘positive infected contacts’. If there are more ‘positive infected contacts’ (e.g. 75) linked to ‘positive spreaders’ (e.g.60) then some of the latter will have been ‘super-spreaders’ (e.g. up to (75-60)/75 = 0.2). The proportion of ‘positive spreaders’ infecting one or more would thus be 0.8, the number being 75*0.8 = 60.

The total number of ‘positive infected contacts’ (e.g. 75) is subtracted from the overall number of newly infected contacts at day 10 (e.g. 275) to give the total number of ‘negative infected contacts’ assumed to have been infected by those originally testing negative at day 0 (e.g. 275=75 = 200). The proportion of super-spreaders infecting these ‘negative infected contacts’ is assumed to be the same as for the ‘positive infected contacts’ (e.g. 0.2). The numbers of negative super-spreaders would therefore be estimated to be 200*0.2 = 40 and the number of ‘negative spreaders’ would be 200-40 = 160.

The reasons for estimating the number of viral spreaders is in order to provide meaningful estimates of the sensitivity, specificity, predictiveness etc of the RT-PCR and LFD tests. However, the efficacy of isolation in terms of risk ratio and the effectiveness of isolation based on RT-PCR and LFD testing can be estimated from the numbers of infected contacts alone. The ratio of spreaders over infected contacts (e.g. 0.8) is the same for the positive and negative spreaders and infected contacts is assumed to be the same in all 3 groups and therefore has no bearing on the estimates of efficacy and effectiveness.

4.2 Simulated results from TT&I

The example ‘observed numbers’ per 100,000 used for the simulation of RT-PCR and LFD results are shown in Table 3.

Table 3: Estimated observed and unobserved numbers of Covid-19 in viral recipients in those isolated and not isolated

| ***x = (b-d)/(c-a) =*** | ***(60-30) / (280-160)*** | ***= 30/120*** | **= 0.25** |
|---|---|---|---|
| OBSERVED number of spreaders per 100,000 in those RT-**PCR test negative** and thus were actually allocated to **NO ISOLATION** | *CALCULATED number of spreaders per 100,000 from RR=0.25 in those RT-**PCR test negative** & imagined allocated to **ISOLATION*** | *CALCULATED number of spreaders per 100,000 from RR=0.25 in those RT-**PCR test positive** imagined allocated to **NO ISOLATION*** | OBSERVED number of spreaders per 100,000in those *RT-**PCR*** **test positive** and thus were actually allocated to **ISOLATION** |
| **a = 160** | *a*r = 160 x 0.25 = 40* | *b/r = 60 / 0.25 = 240* | **b = 60** |
| | | | |
| OBSERVED number of spreaders per 100,000 in those **LFD test negative** and thus were actually allocated to **NO ISOLATION** | *CALCULATED number of spreaders per 100,000 from RR=0.25 in those **LFD test negative** & imagined allocated to **ISOLATION*** | *CALCULATED number of spreaders per 100,000 from RR=0.25 in those **LFD test positive** & imagined allocated to **NO ISOLATION*** | OBSERVED number of spreaders per 100,000 in those **LFD test positive** and thus were actually allocated to **ISOLATION** |
| **c = 280** | *c*r = 280 * 0.25 = 70* | *d/r = 30 / 0.25 = 120* | **d = 30** |

With the results of a = 160, b = 60, c = 280, d = 30, the estimated risk ratio (RR) from Equation 9 is: (d-b)/(a-c) = (30-60)/(160-280)= 0.25. The 'calculated' numbers in Table 3 tell us that the overall proportion of Covid-19 spreaders without isolation is (160+240)/100,000 = 400/100000 = 0.004. The sensitivity of the RT-PCR test is 240/(240+160) = 240/400 = 0.6. As we would know the number of RT-PCRs testing positive (e.g. 343 out of 100,000), the specificity can be calculated from the data in the P Map of Figure 3 where ' A-y-P-x-B' represents 'given A, a proportion of x have B' and 'given B, a proportion of Y have A'. The presence of an arrow (e.g. ' A-y-P-x->B' indicates that A also has a causal effect on B.

The overall proportion with no viral spread is: (100000-400)/100000 = 99600/100000. The proportion with no viral spread conditional on a negative PCR is (99657-160)/99657= 99497/99657. The proportion overall with a negative PCR = (100000-343)/100000 = 99657/100000. From Bayes rule in Figure 1, the specificity is therefore (99657/100000)*(99497/100000)/(99600/100000) = 99497/99600 = 0.998966. The probability of viral transmission conditional on a positive RT-PCR without isolation is 240/343 = 0.7.

Figure 3: A P map of PCR positive / negative & viral spread /no spread with NO targeted isolation

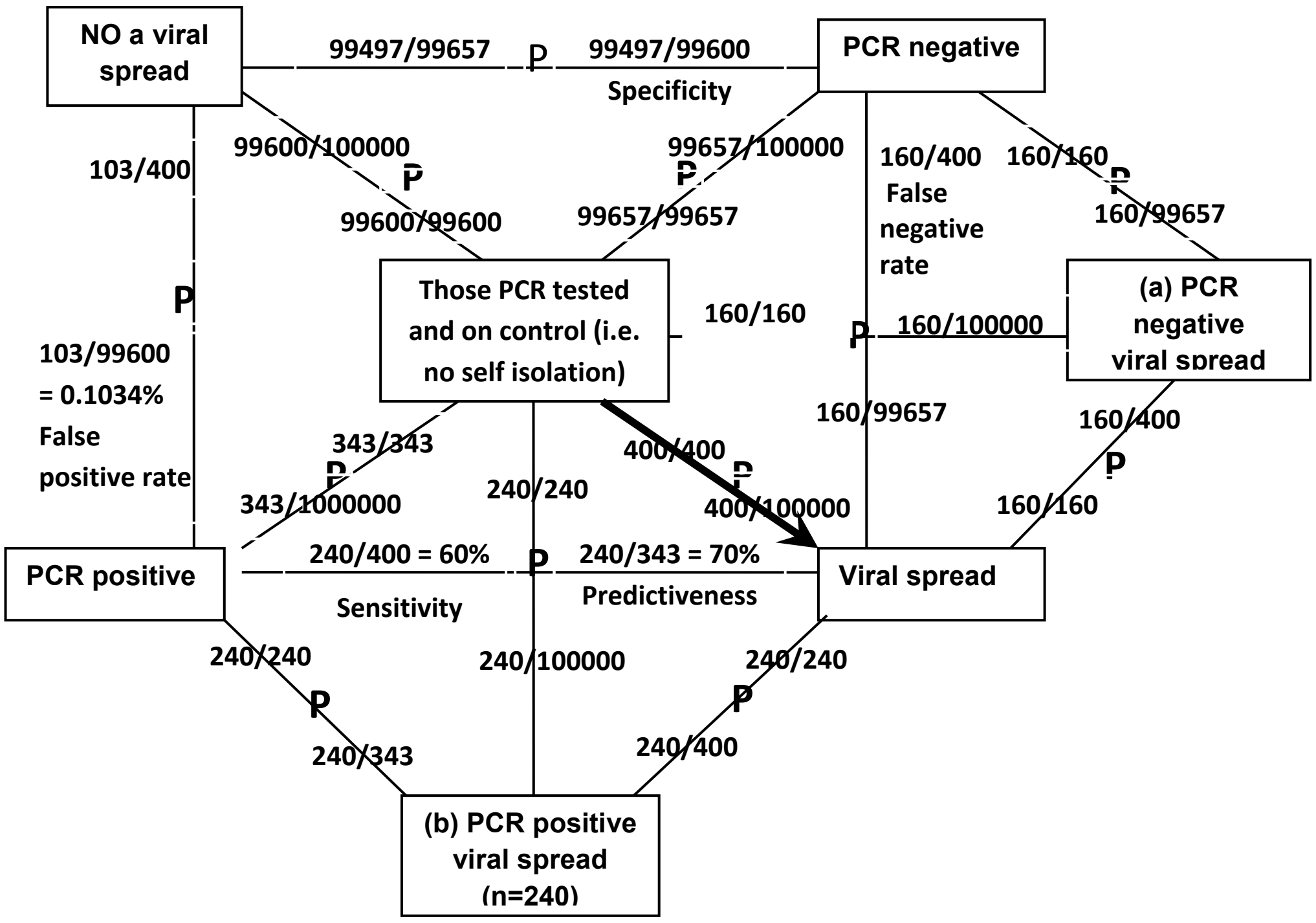

4.3 <u>Discussion of simulation</u>

This simulation shows that if no isolation were done then out of 100,000 subjects, 160+240 or 280+120 = 400 out of 100,000 would have resulted in transmission to at least one other individual. By isolating all those testing RT-PCR positive, 240-60 = 180 fewer or 400-180 = 220 out of 100,000 (instead of 400 out of 100,000) would have resulted in transmission to at least one other individual. However by applying TT&I using LFD, 120-30 = 90 fewer or 310 out of 100,000 (instead of 400) would have resulted in transmission to at least one other individual. However, if in a third trial limb, when isolation occurred more rapidly as soon as the LFD result was known, only 10 would be found to have been spreaders (because the risk ratio was 0.25 *5/15 = 0.0833). This would mean that 120-10 = 110 fewer spreaders would have occurred or 400=110 = 290 spreaders out of 100,000.

The sensitivity of the LFD test from Table 4 is 120/(280+120) = 0.3. As we would know the proportion of LFDs testing positive without isolation (e.g. 133 out of 100,000) and using the same reasoning with proportions as set out in the P Map in Figure 4, its specificity is 99587/99600 = 0.99987. The probability of Covid-19 transmission conditional on a positive LFD would be 120/133 = 0.9. As the risk ratio is 0.25, the probability of viral spread conditional on a positive LFD WITH isolation is 0.9*0.25 = 0.225. The probability of 'benefit' conditional on a positive LFD with isolation is therefore 0.9-0.225 = 0.675. This means that the 0.675 probability of benefit conditional on a positive LFD is greater than 0.525 probability of benefit conditional on a positive RT-PCR. However, fewer people would have a positive LFD (133/100,000) that would have a positive RT-PCR (343 out of 100,000). Therefore, the total number of people benefiting with a positive LFD (133 * 0.9 = 120 out of 100,000) is fewer than the total number benefitting with a positive PCR (343 * 0.7 = 240 out of 100,000).

Figure 4: A P map of LFD positive / negative & viral spread /no spread with NO targeted isolation

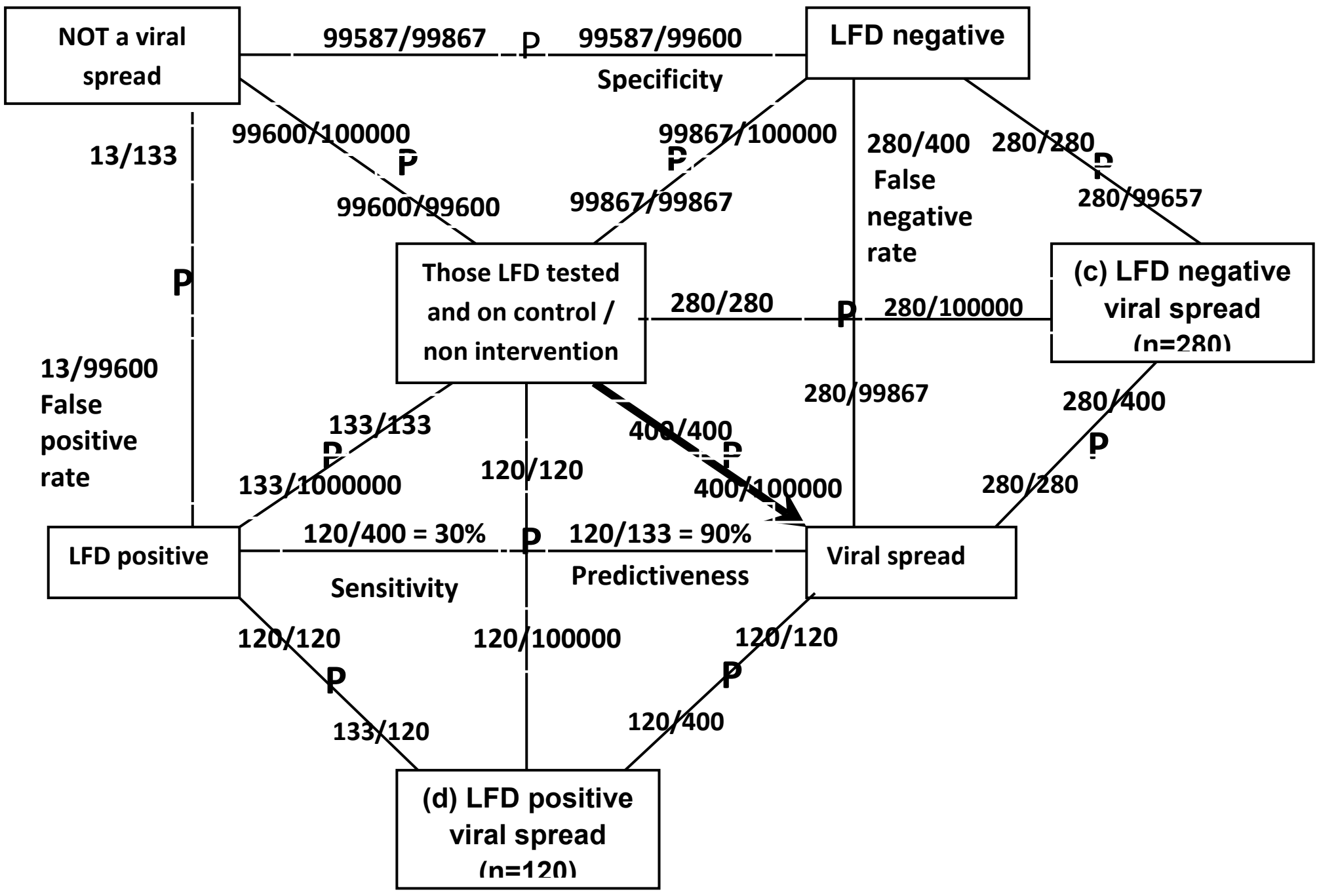

The superiority of the TT&I based on RT-PCR in this simulation is down to its greater assumed sensitivity of 0.6 compared to an assumed sensitivity of 0.3 of the LFD test. This is despite the probability of transmission conditional on a positive LFD (0.9) being higher than that for a RT-PCR (0.7). If a decision to isolate occurred only when both the LFD and RT-PCR tests were positive, then at best this combination would have a sensitivity of 0.3 so that the number of spreaders in those isolated would not change. However, if there was statistical independence between the likelihood of a positive RT-PCR and LFD results, the sensitivity of the combination would be 0.6*0.3 = 0.18. In this case the number of spreaders in those not isolated who were both LFD and RT-PCR positive would be lower at 18 so that with isolation of both LFT and PCR positive people, there would be 72-18 = 54 fewer spreaders. There would therefore be 400-54 = 346 spreaders instead of 400 out of 100,000. Thus isolating only those both LFD and RT-PCR positive would give the worst result. These results are summarised in Table 4 (The Appendix shows other possible results: A1. A result if isolation was ineffective. A2: A result if TT&I were highly effective. A3: The result of using different LFD strategies when isolation is highly effective.

Table 4: Effectiveness of different testing strategies for TT&I

| No TT&I | RT-PCR | LFD + delay | PCR & LFD + delay | LFD no delay |
|---|---|---|---|---|
| 400 | 220 spreaders | 310 spreaders | 346 spreaders | 290 spreaders |
| No fewer | 180 fewer | 90 fewer | 54 fewer | 110 fewer |

5.4 Requirements to make the above simultaneous equations solvable

By determining the numbers of spreaders carefully, it is possible to estimate the performance of T, T & I. In order to be solvable, the simultaneous equations must be mathematically independent. This depends on the tests used being different in terms of their mathematical characteristics such as sensitivity, specificity or predictiveness with respect to 'viral spread'. It must be emphasised that the predictiveness (e.g. of 90 or 70%) of these tests applies to 'spread' and not to diagnosis. These tests are assumed by convention to be sufficient criteria for the diagnosis of Covid-19 and therefore have 100% predictiveness by circular argument. However, they are not definitive because although their positive tests are assumed to identify only those with Covid-19 (because they are assumed by circular reasoning to be 100% specific), they do not identify all those with Covid-19 (because they are not also assumed by the same circular reasoning to be 100% sensitive).

5.1. Modelling conditional probabilities

Instead of setting up simultaneous equations using a pair of different tests such as RT-PCR and LFD, it has already been shown using the AER that this can be done using a pair of different thresholds of a single test. The same principle can also be applied to the RT-PCR test by using two different Cycle thresholds (Ct) to report the result as positive or negative. For example, a positive RT-PCR T1 might be based on a Ct threshold above 25 cycles and a positive RT-PCR-T2 based on a Ct threshold above 35 cycles. The availability of these numerical results can also be used to estimate the probability of spread conditional on individual Ct threshold results by creating conditional probability curves based on ORs or RRs. However, the choice of ORs or RRs for modelling may depends on the collapsibility of ORs or RRs regarding RT-PCR results with respect to the outcome of SARS-Cov-2 virus spread to others or the diagnosis of Covid-19 infection in an individual. Therefore a decision may have to be made to model these conditional probabilities by using calculations based on RRs or ORs

The choice would have to be based on the causal theories applicable to the conditional finding as suggested by Pearl et al [7]. For example, in the case of isolation of Covid-19 sufferers, reduced viral spread to contacts due to isolation of potential viral spreaders would not result in a reduced proportion of PCR positive results in the potential spreaders. This would suggest that the best model should be based on a RR. However, reduction in a leaky kidney capsule from an angiotensin receptor blocker would result in a lower AER. This might suggest that the best model should be based on the OR. These options are discussed next.

5.2. The conditions for collapsibility for modelling with risk ratios and odds ratio

The sets $X_i$=0 (when 'i' = 1 to n) are the 'n' exchangeable sets (e.g. formed by randomisation) prior to being subjected to the effects of a control substance (e.g. placebo) or the various n active substances (e.g. different doses of medication). Thus $X_1$=0 is an exchangeable set destined to be allocated to placebo and become X=1, $X_2$=0 is destined to be allocated an active substance (e.g. a low dose of the drug to become X=2 and $X_3$=0 is destined to be allocated to another substance (e.g. a high dose of the drug) and become X=3. Y=1 is a positive outcome and Y=0 is a negative outcome. $p(Y_{x=1}=1|X=1)$ is the proportion with a positive outcome on control and $p(Y_{x=1}=0|X=1)$ is its complementary negative outcome on control. Z=1 is a positive conditional finding. The proportion with Z=1 before initiating control or intervention is $p(Z_{x=0}=1|X=0)$ and known as a 'baseline' value. The proportion with Z=1 after initiating control is $p(Z_{x=1}=1|X=1)$ and after initiating treatment it is $p(Z_{x=2}=1|X=2)$ or $p(Z_{x>2}=1|X>2)$.

$p(Y_{x=2}=1|X=2)$ is the proportion with a positive outcome in set X=2 after intervention X=2. $p(Z_{x=1}=1|X=1)$ is the proportion with a positive conditional finding Z=1 in set X=1 and $p(Z_{x=2}=1|X=2)$ is the proportion with a positive finding Z=1 after intervention. Both may be different to $p(Z_{x=0}=1|X=0)$ in the exchangeable set. In Figure 3, the set of those not isolated is represented by X=1, a positive RT-PCR is represented by Z=1, the spread of virus to another person is represented by Y=1.

5.3 Collapsibility of marginal risk ratios

Assumption (A) in Section 2.1 was that that the effect measure risk ratio 'r' is caused by the interventions X=1 and X=2 such that

$p(Y_{x=2}=1|X=2) / p(Y_{x=1}=1|X=1) = r$

Assumption (B) in Section 2.1 was that Equation 5.1

$p(Z_{x=0}=1|Y_{x=0}=1) = p(Z_{x=1}=1|Y_{x=1}=1) = p(Z_{x=2}=1|Y_{x=2}=1)$ Equation 5.2

Assumption (B) applies also for Z=0 so that

$p(Z_{x=0}=0|Y_{x=0}=1) = p(Z_{x=1}=0|Y_{x=1}=1)\ p(Z_{x=2}=0|Y_{x=2}=1)$ Equation 5.3

From Bayes rule:

$p(Y_{x=2}=1|X=2)^*\ p(Z_{x=2}=1|Y_{x=2}=1) = p(Y_{x=2}=1\cap Z_{x=2}=0|X=2)$ Equation 5.4

$p(Y_{x=1}=1|X=1)^*\ p(Z_{x=1}=1|Y_{x=1}=1) = p(Y_{x=1}=1\cap Z_{x=1}=0|X=1)$ Equation 5.5

So that

$p(Y_{x=2}=1|X=2)^*\ p(Z_{x=2}=1|Y_{x=2}=1) / p(Y_{x=1}=1|X=1)^*\ p(Z_{x=1}=1|Y_{x=1}=1) = p(Y_{x=2}=1\cap Z_{x=2}=0|X=2) / p(Y_{x=1}=1\cap Z_{x=1}=0|X=1)$

Equation 5.6

But from Assumption (B) in Equation 5.2 that $p(Z_{x=1}=1|Y_{x=1}=1) = p(Z_{x=2}=1|Y_{x=2}=1)$ these values cancel out in Equation 5.6, so that simplifying Equation 5.6 gives:

$p(Y_{x=2}=1|X=2) / p(Y_{x=1}=1|X=1) = r = p(Y_{x=2}=1\cap Z_{x=2}=0|X=2) / p(Y_{x=1}=1\cap Z_{x=1}=0|X=1)$ Equation 5.7

Replacing $p(Y_{x=2}=1|X=2)$ by $p(Y_{x=2}=0|X=2)$ and replacing $p(Y_{x=1}=1|X=1)$ by $p(Y_{x=1}=0|X=1)$ and repeating the reasoning in Equations 5.3 to 5.7:

$p(Y_{x=2}=0|X=2) / p(Y_{x=1}=0|X=1) = r = p(Y_{x=2}=1\cap Z_{x=2}=0|X=2) / p(Y_{x=1}=1\cap Z_{x=1}=0|X=1)$ Equation 5.8

Therefore the marginal probability ratio

$p(Y_{x=2}=0|X=2) / p(Y_{x=1}=0|X=1) = r$ Equation 5.9

and the marginal probability ratio

$p(Y_{x=2}=1\cap Z_{x=2}=0|X=2) / p(Y_{x=1}=1\cap Z_{x=1}=0|X=1) = r$ Equation 5.10

and the marginal probability ratio

$p(Y_{x=2}=1\cap Z_{x=2}=0|X=2) / p(Y_{x=1}=1\cap Z_{x=1}=0|X=1) = r$ Equation 5.11

Therefore as that the marginal probabilities ratios in Equations 5.9, 5.10 and 5.11 are all equal to r, the marginal probabilities are collapsible according to Assumptions (A) and (B).

5.4 Collapsible conditional odds ratios

The odds ratio models a causal connection between the positive outcome Y=1, the negative outcome Y=0, the conditional positive finding Z=1 and the negative finding Z=0. Thus the effect of treatment is to reduce the proportion of patients with the outcome $Y_{x=1}=1$ (e.g. nephropathy on placebo) from $p(Y_{x=1}=1|X=1)$ in those in the control set to $p(Y_{x=2}=1|X=2)$ for those in the treatment set. However, the causal connection between the outcome (e.g. nephropathy) and the finding (e.g. an AER>80mg/min) in those who still have the disease after treatment is assumed to be the same as in those who had the disease without treatment so that the proportion of those with the conditional finding (e.g. an AER >80mcg/min) 'caused' by those with nephropathy is the same in the treated and untreated set. In other words $p(Z_{x=2}=1|Y_{x=2}=1) = p(Z_{x=1}=1|Y_{x=1}=1)$.

The causal connection between those without the disease outcome (i.e. Y=0) and the finding Z=1 may be different (e.g. weaker) but it is also maintained so that $p(Z_{x=2}=1|Y_{x=2}=0) = p(Z_{x=1}=1|Y_{x=1}=0)$. It follows that the likelihood ratio $p(Z_{x=2}=1|Y_{x=2}=0)/p(Z_{x=2}=1|Y_{x=2}=1) = l$ and the likelihood ratio $p(Z_{x=1}=1|Y_{x=1}=0)/p(Z_{x=1}=1|Y_{x=1}=1) = l$ are the same.
This means that when 'q' is the odds ratio
$odds(Y_{x=2}=1|X=2) / odds(Y_{x=1}=1|X=1) = q$ Equation 5.12
and
$odds(Y_{x=2}=1|Z_{x=2}=1) / odds(Y_{x=1}=1|Z_{x=1}=1) = q= odds(Y_{x=2}=1|X=2)*l / odds(Y_{x=1}=1|X=1)*l =q$
Equation 5.13
Cancelling out the 'l' on the right hand side of Equation 5.13:
$odds(Y_{x=2}=1|Z_{x=2}=1) / odds(Y_{x=1}=1|Z_{x=1}=1) = q = odds(Y_{x=2}=1|X=2) / odds(Y_{x=1}=1|X=1) = q$
Equation 5.14
By the same reasoning in equations 5.13 and 5.14:
$odds(Y_{x=2}=1|Z_{x=2}=0) / odds(Y_{x=1}=1|Z_{x=1}=0) = odds(Y_{x=2}=1|X=2) / odds(Y_{x=1}=1|X=1) = OR$ Equation 5.15
The three odds ratios in Equations 5.12, 5.14 and 5.15 are the same. Therefore the odds ratios are collapsible if the likelihood ratios are the same in the intervention and control sets. This implies that there is a causal effect from the outcomes to the conditional findings.

Note that $p(Z_{x=2}=1|X=2)$ is the new probability of the previously baseline value of $p(Z_{x=0}=1|X=0)$ after treatment X=2 has been started (e.g. a few days after starting an angiotensin receptor blocker irbesartan and the AER has settled to a new level). If this measurement has not been made and is theoretical, this can be signified by p' as above instead of p. The observed probability of pre-treatment level is indicated by $p(Z_{x=0}=1|X=0)$. In practice, it is assumed that the probabilities of the conditional finding Z=1 conditional to being on placebo are the same as those contional on the exchangeable set before placebo is started so that $p(Z_{x=1}=1|X=1) = p(Z_{x=0}=1|X=0)$. If it can be assumed that there is a causal connection between both Y=1 and Z=1 and Y=0 and Z=1, then it can be assumed that there is constant likelihood ratio between sets X=1 and X=2 (i.e. that $p(Z_{x=2}=1|Y_{x=2}=0)/p(Z_{x=2}=1|Y_{x=2}=1) = p(Z_{x=1}=1|Y_{x=1}=0)/p(Z_{x=1}=1|Y_{x=1}=1)$. This will then mean that if $p(Y_{x=2}=1|X=2) < p(Y_{x=1}=1|X=1)$ then $p(Z_{x=2}=1|X=2) < p(Z_{x=1}=1|X=1)$.

5.6 Collapsibility of conditional risk ratios

The situation for collapsibility of conditional risk ratios is different in that must be no causal effect of the outcome on the conditional finding. This is satisfied for example by reduced spread due to self-isolation having no effect on the original proportion with a positive PCR after isolation starts by reducing the proportion with a positive PCR. In this situation, when the proportion with a positive PCR before isolation is $p(Z_{x=1}=1|X=1)$ and the proportion after isolation is $p(Z_{x=2}=1|X=2)$ then $p(Z_{x=2}=1|X=1) = p(Z_{x=2}=1|X=1) = p(Z_{x=0}=1|X=0)$, which is different for odds ratio.. Therefore when $p(Z_{x=2}=1|X=2) = p(Z_{x=2}=1|X=1)$ and $p(Z_{x=2}=1|Y_{x=2}=1)/p(Z_{x=1}=1|Y_{x=1}=1)$ and that if

$p(Y_{x=2}=1|X=2) / p(Y_{x=1}=1|X=1) = r$ Equation 5.16

from Bayes rule

$$\frac{p(Y_{x=2}=1|p(Z_{x=2}=1)}{p(Y_{x=1}=1|p(Z_{x=1}=1)} = \frac{p(Y_{x=2}=1|X=1)*p(Z_{x=2}=1|Y_{x=2}=1) / p(Z_{x=2}=1|X=2)}{p(Y_{x=1}=1|X=1)*p(Z_{x=1}=1|Y_{x=1}=1) / p(Z_{x=1}=1|X=1)} = r \qquad \text{Equation 5.17}$$

The shaded values cancel out so that

$p(Y_{x=2}=1|p(Z_{x=2}=1) / p(Y_{x=1}=1|p(Z_{x=1}=1) = r$ Equation 5.18

and by the same argument

$p(Y_{x=2}=1|p(Z_{x=2}=0) / p(Y_{x=1}=1|p(Z_{x=1}=0) = r$ Equation 5.19

This confirms that conditional risk ratios are collapsible when $p(Z_{x=2}=1|X=2) = p(Z_{x=2}=1|X=1)$ and $p(Z_{x=2}=1|Y_{x=2}=1)/p(Z_{x=1}=1|Y_{x=1}=1)$ (i.e. when the overall proportions with the conditional finding are the same in the control and treatment sets and the likelihoods of the conditional finding conditional on the outcome are also the same in the control and treatment sets). This implies that there is no causal connection from the outcome to the conditional finding.

5.7 The theoretical nature of strict conditional collapsibility

The precise conditions of probabilities of findings being equivalent in the control and intervention set for conditional RRs and the likelihood ratios being equivalent for ORs to be collapsible can only be confirmed or refuted after the true probabilities are known after an infinite number of observations. They are therefore theoretical conditions for use in mathematical modelling. The choice of model based on constant odds ratio or risk ratio would depend on the causal considerations discussed in sections 5.4 and 5.5. However, both models were used to create the curves in Figure 5.

The likelihood distributions for those with and without the outcome were formed by pooling the data from those on placebo and treatment. This ensured that when Z=i was a particular AER value in mcg/min, when X=1 was placebo and X=2 was treatment, Y=1 was nephropathy and Y=0 was no nephropathy, then $p(Z_{x=1}=i|Y_{x=1}=1) = p(Z_{=2}=i|Y_{x=2}=1)$ and $p(Z_{x=1}=i|Y_{x=1}=0) = p(Z_{=2}=i|Y_{x=2}=0)$ ensuring that the odds ratio between $p(Y_{x=1}=i|Z_{x=1}=1)$ and $p(Y_{=2}=i|Z_{x=2}=1)$ for all values of Z=I were collapsible and therefore constant. The curve based on a constant risk ratio also used the distributions $p(Z_{x=1}=i|Y_{x=1}=1) = p(Z_{=2}=i|Y_{x=2}=1)$. However in order that $p(Z_{x=1}=i|X=1) = p(Z_{x=1}=i|X=2)$ a spline distribution was fitted to all the AER data in the study so that when this one distribution was used to calculate $p(Y_{x=1}=i|Z_{x=1}=1)$ and $p(Y_{=2}=i|Z_{x=2}=1)$, it ensured that the risk ratio between all $p(Y_{x=1}=i|Z_{x=1}=1)$ and $p(Y_{=2}=i|Z_{x=2}=1)$ for all values of Z=i were collapsible and therefore constant.

Figure 5 shows curves that display the probabilities of nephropathy on treatment conditional on the AER. The top pair displays the probabilities of nephropathy on placebo. The unbroken curve is based on fitting likelihood distributions with Gaussian kernel splines to the data and calculating the probabilities using Bayes rule. The SD of the kernels was set to give probability confidence limits of +/- 0.05 at an AER of 80mcg/min. The broken curve is the result of calibrating the calculated curve by adjusting a calibration constant (when The calibrated probability = (The uncalibrated probability)*(1+(the AER-80)*Adjustable calibration constant)) until (1) the average of the probabilities from each of the 173 data points on the placebo curve above 80mcg/min was equal to the proportion of 20/62 from Table 1 and (2) the average of the probabilities of nephropathy 173 data points above 80mcg/min on both treatment curves was equal to 19/112 from Table 1. The curves above an AER of 80mcg/min were given priority as it is here is where the discrepancies of greatest value would be found. The marked difference between the calculated and calibrated placebo curves in the upper AER range reflects sparseness of data there, the estimated 95% confidence limits being +/- 0.3 based on the estimated number of data points contributing to the spline curve's likelihood probability density at 200mcg/min.

Figure 5: Curves displaying the probability of nephropathy conditional on the individual patient's albumin excretion rate based on odds ratio and risk ratio with and without calibration

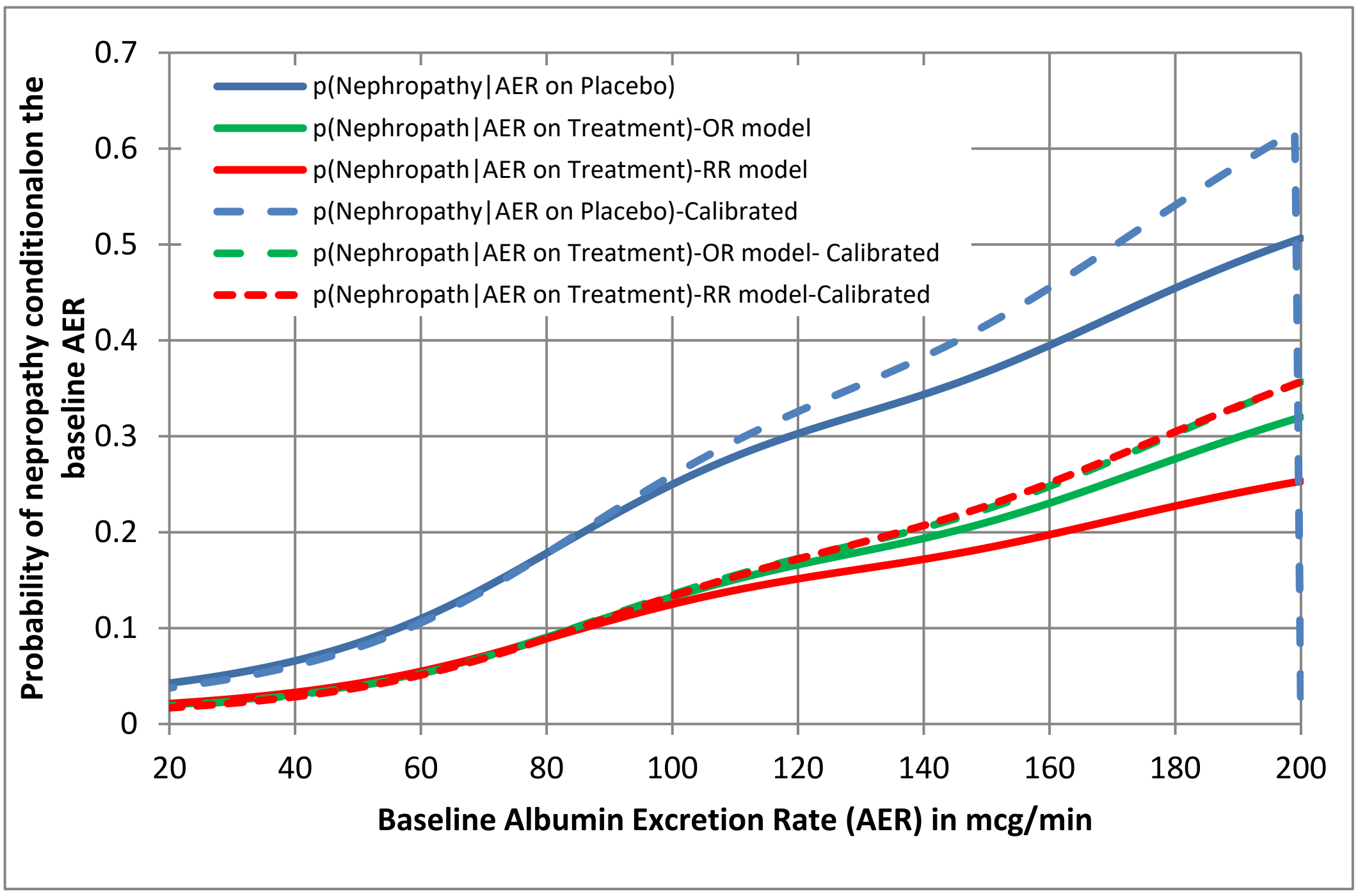

The bottom curve displays the probabilities of nephropathy on treatment based on a constant risk ratio. The curve displaying the probabilities of nephropathy on treatment based on a constant odds ratio is the 2nd from the bottom. The broken curves show the calibrated probabilities calculated in the same way as for the placebo curve. It can be seen that the calibrated curves for the odds ratio and risk ratio models are superimposed, which is what one would expect. The difference between the theoretical and calibrated odds ratio curves is small. However the difference is greater between the theoretical and calibrated risk ratio curves. Therefore when both curves based on theory were calibrated against the sparse data obtained so far, less calibration was required for the curve based on the odds ratio, so the data so far leans more towards the odds ratio model. The unbroken curves are what might be expected theoretically after an infinite amount of data was collected. This is in keeping with the discussion in sections 5.5 and 5.6 that suggested that the most appropriate model should be based on the odds ratio due to the causal effect of treatment on the AER.

## 6. Conclusion

It is possible to estimate the overall risk ratio of in the outcomes of a clinical trial by randomising subjects to two different testing strategies instead of randomizing them directly to an intervention or control. When the outcome on control (e.g. nephropathy as indicated by heavy proteinuria on placebo or a contact converting from RT-PCR or LFD negative to positive) is regarded as the outcome, it is possible to assess an individual test result's ability to predict this outcome. This also would give the test's positive predictiveness, sensitivity and specificity regarding the adverse outcome (not diagnosis). The assumption of equivalent likelihood distributions and constant odds ratios could be used to create model curves that after calibration against the latest data would display provisional probabilities of the outcome on intervention and control to be updated as new data become available.

## Acknowledgments
I am grateful for the support of the past employees of Sanofi-Synthelabo and Bristol-Myers Squibb, and the investigators in numerous countries who participated in the IRMA2 trial for providing the data that helped me to develop these methods.

APPENDIX

A.1 A result if isolation was ineffective

If the following observations in Table 5 were made, this would indicate that isolation was ineffective with a risk ratio of 1 but the performance of the PCR and LFT tests were the same as in Table 3. The same result could be obtained by performing the RT-PCR and LFD tests on the same patients, controversially (i.e. unethically) advising those testing both positive and negative for LFD and RT-PCR not to isolate at all and then observing the proportion of patients who went on to transmit to contacts of the positive and negative groups for both tests.

Table A1: Simulated data that suggest completely ineffective isolation

| OBSERVED number of spreaders per 100,000 in those RT-PCR test negative and thus were actually allocated to NO ISOLATION | *CALCULATED number of spreaders per 100,000 from RR= 1 in those RT-PCR test negative & imagined allocated to ISOLATION* | *CALCULATED number of spreaders per 100,000 from RR=1 in those RT-PCR test negative imagined allocated to NO ISOLATION* | OBSERVED number of spreaders per 100,000 in those *RT-PCR* test positive and thus were actually allocated to ISOLATION |
|---|---|---|---|
| 160 | *160 x 1 = 160* | *240/1 = 240* | 240 |
| | | | |
| OBSERVED number of spreaders per 100,000 in those LFD test negative and thus were actually allocated to NO ISOLATION | *CALCULATED number of spreaders per 100,000 from RR=1 in those LFD test negative & imagined allocated to ISOLATION* | *CALCULATED number of spreaders per 100,000 from RR=1 in those LFD test negative & imagined allocated to NO ISOLATION* | OBSERVED number of spreaders per 100,000 in those LFD test positive and thus were actually allocated to ISOLATION |
| 280 | *280 x 1 = 280* | 120 / 1 = 10 | 120 |

If the PCR and LFD tests were both useless because their sensitivities and false positive rates were the same and there was no risk ratio (i.e. the risk ratio was 1), then all four observed outcomes and four calculated outcomes would be the same. If the following observations in Table 6 were made, this would indicate that both LFD and RT-PCR were highly predictive and that isolation highly effective so that there was a major impact on reducing transmission.

A.2 An example result if TT&I were highly effective

Table 6 tells us that the sensitivity of the RT-PCR test is 120/(120+80) = 0.6. As we know that the observed PCR positive tests was 343 out of 100,000, its specificity is (50000*((100000-300)/100000)-120+80)/(50000-120) = 0.998597.

Table A2: Simulated data that suggest highly effective TT&I

| OBSERVED number of spreaders per 100,000 in those RT-PCR test negative and thus were actually allocated to NO ISOLATION | *CALCULATED number of spreaders per 100,000 from RR=0.1 in those RT-PCR test negative & imagined allocated to ISOLATION* | *CALCULATED number of spreaders per 100,000 from RR=0.1 in those RT-PCR test negative imagined allocated to NO ISOLATION* | OBSERVED number of spreaders per 100,000 in those *RT-PCR* test positive and thus were actually allocated to ISOLATION |
|---|---|---|---|
| 80 | *8 x 0.1 = 8* | 12 / 0.1 = 120 | 12 |
| | | | |
| OBSERVED number of | *CALCULATED number* | *CALCULATED number of* | OBSERVED number of |
| spreaders per 100,000 in those LFD test negative and thus were actually allocated to NO ISOLATION | *of spreaders per 100,000 from RR=0.1 in those LFD test negative & imagined allocated to ISOLATION* | *spreaders per 100,000 from RR=0.1 in those LFD test negative & imagined allocated to NO ISOLATION* | spreaders per 100,000 in those LFD test positive and thus were actually allocated to ISOLATION |
| 40 | *4 x 0.1 = 5* | 16 / 0.1 = 160 | 16 |

The sensitivity of the LFD test from Table 6 is 60/(160+40) = 0.8. As we know that the observed LFD positive tests was 133 out of 100,000, its specificity is (50000*((100000-323)/100000)-160+40)/(50000-160) = 0.997562.

Table A3 shows the result of using different LFD strategies when isolation is highly effective.

Table A3: the number of spreaders per 100,000 after different testing strategies for TT&I

| No TT&I | RT-PCR | LFD + delay | LFD no delay |
|---|---|---|---|
| 400 spreaders | 184 spreaders | 112 spreaders | 96 spreaders |
| No fewer | 216 fewer | 288 fewer | 304 fewer |